\documentclass[runningheads]{llncs}
\usepackage{graphicx}
\usepackage{wasysym}
\usepackage{colortbl}
\usepackage{soul}
\begin{document}
\title{Exploring Virtual Reality through Ihde’s Instrumental Realism}
%
%
\author{He Zhang\orcidID{0000-0002-8169-1653} \and
John M. Carroll\orcidID{0000-0001-5189-337X}}
\authorrunning{H. Zhang and J M. Carroll}
\institute{Pennsylvania State University, University Park, PA 16802, USA \\
\email{hpz5211@psu.edu};
\email{jmc56@psu.edu}}
\maketitle              
\begin{abstract}
Based on Ihde's theory, this paper explores the relationship between virtual reality (VR) as an instrument and phenomenology. It reviews the ``technological revolution'' spurred by the development of VR technology and discusses how VR has been used to study subjective experience, explore perception and embodiment, enhance empathy and perspective, and investigate altered states of consciousness. The paper emphasizes the role of VR as an instrumental technology, particularly its ability to expand human perception and cognition. Reflecting on this in conjunction with the work of Husserl and Ihde, among others, it revisits the potential of VR to provide new avenues for scientific inquiry and experience and to transform our understanding of the world through VR.

\end{abstract}
\section{Introduction}

Virtual Reality (VR) was invented by Ivan Edward Sutherland in 1965~\cite{stark2022major}, and the technology is considered a window into the virtual world. This technology allows users to immerse themselves in a (virtual) world through a head-mounted display (HMD). VR has evolved from a lab product to a consumer-grade market over the decades and has become a widely popular technology. The high presence (immersion or realism) and ability to break the boundaries of reality that VR environments possess have received widespread popularity. In the industrial world, many companies have been involved in the development of VR devices and applications, the most representative ones being Meta (Oculus), HTC, Sony, Pico, etc. According to Market Research Report 2022~\cite{vrmarketreport2022}, the global VR market is expected to grow from \$16.67 billion in 2022 to around \$227.34 billion in 2029. VR applications show extraordinary promise in various fields~\cite{kim2016virtual}, including but not limited to entertainment~\cite{bates1992virtual}, education~\cite{rojas2023systematic}, healthcare~\cite{halbig2022opportunities}, and creativity~\cite{wiseman2022experiencing}.

In this article, we will focus on the content and theory of Ihde's book~\textit{``Instrumental realism: The interface between philosophy of science and philosophy of technology''}~\cite{ihde1991instrumental} to explore how VR as an instrument extends the boundaries of science. By examining the content and theory of Ihde's work, this paper aims to shed light on the implications of VR for human-computer interaction (HCI), phenomenology, and scientific instruments. Specifically, we will discuss Ihde's instrumental realism and its implications for VR technology. Firstly, we will discuss the role of VR as a technologically revolutionary scientific instrument and introduce the paradigm shift triggered by VR in different fields. Then, we will introduce VR in relation to the phenomenology of perception and further discuss the paradigm shift in the context of interactive that VR technology provokes. Finally, we will discuss how to understand and how to use VR for friendly interaction. 

\section{Technological Revolution}
Advances in VR technology have transformed the way we interact with and understand the world around us. With VR devices, people can create a new space or world with their thought. Similar to how other breakthrough technologies, such as telescopes and microscopes, have changed our perception of reality, VR has helped define the boundaries of our experience through its immersive, create-able capabilities that complement our understanding of the world or self-created world. 

Specifically, VR technology can open up new possibilities for scientific inquiry and experimentation by immersing users in virtual environments to experience or experience rich content that breaks the limits of the real world~\cite{lv2020virtual}.

Ihde~\cite{ihde1991instrumental} discusses Kuhn and Foucault's views on paradigm shifts and scientific revolutions. For Kuhn, a new paradigm represents a rare phenomenon, which he terms a ``revolution''. In contrast, Foucault perceives this as an ongoing process, suggesting that paradigms are in a state of continuous evolution. We refer to the paradigm shift associated with VR as a technological revolution rather than a mere paradigm shift for two reasons. First, even though the concept of VR has been around for decades, it has been ``constructed'' in juxtaposition to the natural sciences; it doesn't simply exist in nature waiting to be observed. Second, VR extends the realm of human behavior. In essence, individuals utilize VR to engage in activities previously unimaginable. VR has already had a major impact on many aspects of real life and has truly changed the paradigm~\cite{heim1993metaphysics}. Here are a few examples.

\textbf{Enhanced communication}: VR allows us to connect with others more immersive and engagingly. It breaks down geographic barriers and allows people to interact with each other in shared virtual spaces, fostering collaboration and social connection. Communication through VR allows attendees to be in the same (virtual) space~\cite{rogers2022realistic} beyond the original concept of smartphones and instant messaging~\cite{ingram2000beyond}. VR technology has changed the paradigm of online communication. Before the advent of VR, our online communication was limited to text-based messages, phone calls, and video chats. While these tools have helped us stay connected, they still lack the immersive and engaging experience like VR. Through VR, we can see and hear others as if they were in the same space and even manipulate virtual objects collaboratively. This level of immersion and interactivity has transformed how we collaborate, learn, and socialize online. It has significant implications for personal relationships and professional teamwork, as VR enables us to communicate and collaborate more naturally, fostering social connections and significantly strengthening players' social closeness in particular~\cite{10.1145/3544548.3581230}.

\textbf{Immersive education and training}: VR is changing the way people learn and develop skills by providing highly interactive and immersive learning experiences. In a virtual environment, learners can better understand the subject matter by interacting ``closer up'' with the content~\cite{takeuchi2011kids,hudson2018using}. At the same time, through VR, learners can continuously practice skills to improve performance in a ``real'', safe, and controlled environment. This approach avoids potential injury to learners during real-world training while effectively reducing costs. Compared to traditional teaching methods, which rely on textbooks, lectures, and memorization in a tedious and unattractive format, VR technology offers a more engaging, interactive, and lower-cost learning method~\cite{haque2006meta,mao2021immersive,aim2016effectiveness}.

\textbf{Entertainment and gaming}: 
The entertainment industry is significantly impacted by VR technology today. On the one hand, the way of participating in games has been changed because of VR~\cite{squire2007wherever}. In the virtual environment, the gaming experience is not even lower than in the real world~\cite{pastel2021comparison}. Instead, because of more flexible controls and a higher degree of freedom in VR, most users' gaming experience has been improved~\cite{10.1145/3332165.3347875}. Taking traditional video games as an example, it usually requires players to stay in front of a monitor with a controller or keyboard~\cite{wolf20021,yuan2011game}. VR does not have any limitations like that way. VR is not only able to let users get into a ``real'' virtual world but also gives a more immersive and interactive experience. As well as, with the development of VR, it has become possible to drop the controller away in the virtual environment completely~\cite{10.1145/3359996.3364240,yang2019gesture,masurovsky2020controller}. On the other hand, with the development and lower cost of VR technology, it has become less expensive to immerse in a game environment for an impressive experience. For example, VR can replace the costly and complex maintenance of escape rooms or flight simulator cockpits. VR has greatly improved the accessibility of content for the general user.

It is fair to say that VR has greatly improved the accessibility of content for the general user and expanded creative expression and entertainment possibilities, providing users with a more engaging and immersive experience.

\textbf{Professional industries}: VR technology has changed the paradigm of how professionals work in multiple industries.  For examples, architects and engineers can use VR to visualize and manipulate their designs in three-dimensional space, enabling better decision-making and problem-solving~\cite{portman2015go,alsafouri2019mobile}. Medical professionals can utilize VR for surgical simulations and therapy~\cite{gallagher2005virtual,javaid2020virtual}, while businesses can leverage VR for marketing, product demonstrations, and employee training~\cite{boyd2019introduction}.

\section{VR as an Instrument in Phenomenology, Perception, and Practice}
In Ihde's book ``Instrumental Realism'', he emphasizes the vital relationship between science and technology, specifically focusing on the role of instruments in scientific practice~\cite{ihde1991instrumental}. Ihde's instrumental realism posits that instruments are not merely passive tools but actively shape and extend human perception and understanding of the world. They are deeply embedded in knowledge production and serve as mediators between humans and their environment. Ihde applied Latour's approach and the concept of ``technoscience'', where science and technology are intertwined and cannot be separated, to recognize the practical (phenomenological) significance of science, the ``sociogical epoche''. 

In this section, we will discuss the role and relationships between VR technology and phenomenology, perception, and practice.

\subsection{VR and Phenomenology}
Phenomenology~\cite{husserl2012ideas} is a philosophy initiated by Edmund Husserl at the beginning of the
twentieth century. Phenomenology, as a presuppositionless science and a philosophical approach, emphasizes the study of human experience and the way things appear in our consciousness, specifically, the human experience of an object or about an object~\cite{moran2002introduction,giorgi2003phenomenology}. 

Ihde highlights the ways in which individuals perceive and engage with the world. In his phenomenological analysis, ``embodiment relations'' is an important concept, i.e., embodiment, hermeneutic, alterity, and background relations. He uses this concept to describe how technology affects our perception and experience of the world, which is referred to in his other book, ``Bodies in Technology''~\cite{ihde2002bodies}.

VR technology can provide a unique platform for phenomenological investigation, allowing researchers to study human experience in a controlled and simulated environment. For example, VR can be used to study how individuals perceive and interpret their surroundings, or to explore the nature of perception, embodiment, and agency in virtual environments. By placing users in immersive, interactive, and often novel environments, VR technologies can help reveal the fundamental structure of human experience, potentially contributing to our understanding of consciousness and subjective reality.

In the context of VR, phenomenology can exist in several ways:

\textbf{Understanding subjective experiences}: VR technology can provide a unique platform for studying the human experience in immersive and interactive environments. By placing users in various ``real'' virtual situations, researchers can study their subjective experiences~\cite{kim2020systematic}, emotions~\cite{diemer2015impact,zhang2023decoding,zhang2024vrmnbd}, and reactions~\cite{peperkorn2016representation,zhang2024vrmnbd} in a controlled environment, which can reveal the fundamental structure of human experience and contribute to our understanding of consciousness and subjective reality~\cite{heater1992being}.

\textbf{Exploring perception and embodiment}: VR can be used to explore the nature of perception, embodiment, and agency in virtual environments. For example, researchers can study how users perceive and interpret their virtual environments, how they feel a sense of presence or immersion~\cite{cummings2016immersive}, and how they interact with virtual objects and characters~\cite{kang2020comparative,hoffman1998physically}. These investigations can provide valuable insights into the interplay of perception, cognition, and action in virtual and real-world contexts.

\textbf{Enhancing empathy and perspective-taking}: VR has been proven to be an effective tool for developing empathy and perspective by allowing users to experience situations from the perspective of others~\cite{van2018virtual}. Through immersive and interactive narratives, VR can help individuals better understand the experiences and emotions of people from different backgrounds and cultures~\cite{georgiadou2021equality}. These studies can contribute to the phenomenological study of empathy and intersubjectivity.

\textbf{Investigating altered states of consciousness}: VR can simulate altered states of consciousness or create environments that challenge our perceptual expectations, such as simulated dreaming~\cite{suzuki2017deep}. Also, VR can provide a way for researchers to study the nature of consciousness and the boundaries of human experience in a controlled and safe manner, such as for the study of hallucinations~\cite{suzuki2018hallucination}.


\subsection{VR and Perception}
\textit{``If the ‘world’ changes in a paradigm shift, the object of reference of perception within its entire field changes: it reflexively implies a change of some kind in the perceiver''}~\cite{ihde1991instrumental}.

When paradigms change, it changes everything about how science views and explains different scenarios. Agian, when the science community discovers something new compared to the original paradigm, a paradigm shift also occurs~\cite{margolis1993paradigms}. For example, the Ancient Greeks' Perception of the Earth was to believe the earth is flat. But once Aristotle first recognized that the earth is a round sphere, a 'paradigm shift' occurred in the scientific community~\cite{vosniadou1992mental}. In the context of VR, Ihde's instrumental realism offers valuable insights into the way VR functions as an instrument that extends human perception and cognition. As a technology that allows users to experience immersive and interactive environments, VR has the potential to transform our understanding of the world by presenting new possibilities for observing, experimenting, and learning. VR environments can provide scientists with new ways of investigating phenomena that were previously inaccessible or difficult to study, thus expanding the boundaries of scientific inquiry~\cite{bainbridge2007scientific}.

The illusionary mechanisms provided by VR allow users to react realistically to VR scenes, effectively providing an immersive experience. The first is the place illusion (PI), also known as ‘being there’ or ‘presence’, which refers to a sense of being in a real place. The second is the plausibility illusion (PSI), an illusion that the events being portrayed are actually happening~\cite{slaterPlaceIllusionPlausibility2009}. VR makes users believe they are in the environment of the game (PI) experiencing the scene as it is happening (PSI)~\cite{linFearVirtualReality2017}. With advances in VR technology, this illusionary mechanism is achieved primarily through head-mounted displays combined with precise motion tracking systems, allowing the user to experience an interactive 3D virtual environment~\cite{davis2009avatars}. One example is that players who play horror games through VR experience a much stronger sense of fear and anxiety than those playing in 2D video mode~\cite{pallaviciniEffectivenessVirtualReality2018}.

\textit{``In Husserl’s case, what is fundamental is a kind of ordinary human praxis and perception, the world of the human interaction among material things and others. Its openness toward the other is sensory, and this relation is focally perceptual''}~\cite{ihde1991instrumental}. 

VR can significantly change human perception by providing an immersive and interactive experience that changes the way we process and interpret information about our surroundings. When people are in a virtual environment, this ``fake'' scenario brings people a real experience.

Also, in Markham's book~\cite{markham1998life}, ``Life Online: Researching Real Experience in Virtual Space'', she discusses how reality is defined or experienced in virtual space and the relationship between the virtual and the ``body''. Markham emphasizes the importance of the ``real'' experiences and emotions that individuals experience in virtual spaces, i.e., the experiences they have in virtual spaces are as important and influential as the interactive experiences they have in the real world. 

In the VR environment, the body interacts in the virtual space and also perceives the world around it, and this virtual space is able to transcend the limitations of the real world effortlessly.


\subsection{VR as an Instrument}
\textit{``Led by a new paradigm, scientists adopt new instruments and look in new places''}~\cite{ihde1991instrumental}.
In fact, VR is an instrumental technology, not only because it is inherently device-dependent (a combination of HMD and sensors) but also because VR is a ``readable technology~\cite{heelan1983natural}'' that allows people to use the instrument to observe and experience the virtual world~\cite{frontoni2018cyber}. 

Moreover, we would like to emphasize the importance of VR, a scientific instrument, as a medium. VR serves as an intermediary between the real and virtual worlds, allowing for tracing features in virtual space and real life. This expands the boundaries of the real world in a sense, while giving extreme freedom. In other words, the restrictions for the virtual world are greatly reduced compared to those in the real world. 

On the one hand, thanks to the immersion, interactivity, scalability, and accessibility that VR offers, conducting scientific research in a virtual environment has practical implications for real life.

On the other hand, in Ihde's view~\cite{ihde1991instrumental}, \textit{```micro-technologies' of early physics were not yet capable of manipulation of natural phenomena in any powerful or significant way. The larger the macrophenomenon and the smaller the instrument, the more limited one is to the mere `observational' model.''} But, VR as an instrument makes it possible to manipulate phenomena in an ``arbitrary'' way in virtual space~\cite{pirok2006making}. At the same time, the empathetic experience of the VR environment allows perception beyond the limits of ``observation''~\cite{el2019virtual}.

In essence, we believe that VR technology has become an advanced scientific instrument and has the potential to grow even more. This is consistent with Ihde's view\cite{ihde1991instrumental}, \textit{``...that contemporary science is more than accidentally-it is essentially-embodied technologically in its instrumentation''}, which emphasizes that contemporary science is fundamentally embodied in its technological tools and that VR is a prime example of this embodiment. As we progress, the combination of VR and scientific exploration will likely deepen, leading to discoveries and innovations that will further expand our understanding of the natural and virtual worlds.

\section{Counterarguments, Counterexamples, and Challenges}

Although people have become aware of the uses and potential possibilities of VR, the VR environment also comes with some risks, especially in terms of its impact on the real world. Firstly, in a highly simulated world, distinguishing between reality and simulation becomes difficult, leading to a loss of meaning in the ``real'' world~\cite{baudrillard1994simulacra}. Secondly, an overemphasis on technology (such as VR) in understanding human experience can lead to a deterministic viewpoint, where technology is seen as shaping everything, thereby undermining human agency~\cite{brey2009philosophy}. Concerns about VR leading to escapism, loss of genuine interpersonal relationships, or issues with privacy and data security challenge the unbridled optimism surrounding VR technology~\cite{rushkoff2022survival,stephanidis2019seven,10130406}.

In the face of these challenges, the use of phenomenology to focus on subjective experience is crucial for designing more immersive, intuitive, and meaningful user-centered VR technologies. This includes paying attention to embodied cognition~\cite{foglia2013embodied} (the interaction between the body and the world), media theory~\cite{mcluhan2017medium} (medium - message), replacing reality~\cite{ryan2001narrative}, and the effects of other technological developments on VR, such as Large Language Models~\cite{10302997,zhang2023redefining}, virtual society~\cite{10098667}, Internet of Things~\cite{10002946}, and digital twins~\cite{zhang2023multichannel}.

\section{Conclusion}
Focusing broadly, our emphasis should remain rooted in the practice of science. VR emerges not merely as a novelty but as an influential instrument that has the potential to redefine our perceptions and interactions with the world, enhancing the very essence of scientific practice.

From a research vantage point, VR can be envisioned as a modern-day extension of our perceptual and cognitive faculties. This opens up fresh avenues for detailed observation, inventive experimentation, and profound learning. These are foundational for the effective implementation of phenomenological methods and understanding. The unique potential of VR to craft, manipulate, and correlate phenomena across the real and simulated realms stands to revolutionize scientific research and, by extension, our daily interactions and experiences.

To conclude, when viewed through the intricate lenses of phenomenology and Ihde's Instrumental Realism, VR emerges as a profound contributor to our comprehension of the human journey. As it blurs the lines between tools and perception, and as phenomenological insights weave new definitions of consciousness, VR finds itself at a pivotal juncture. This juncture teems with prospects of new discoveries, deep self-reflection, and growth. 

%
%
%
 \bibliographystyle{splncs04}
 \bibliography{custom}

\end{document}